\documentclass[aip,rsi,graphicx]{revtex4-1} 
\usepackage{graphicx}
\usepackage[author={mcavedon}]{pdfcomment}
\usepackage{bm}
\usepackage{siunitx}
\usepackage{cleveref}
\usepackage{tikz}
\usetikzlibrary{patterns}
\usepackage{lineno}
\usepackage{amsmath}

\draft 

\begin{document}


\title{A fast acquisition rate system for charge exchange
measurements at the plasma edge at the ASDEX Upgrade tokamak} 



\author{M. Cavedon}
\email{marco.cavedon@ipp.mpg.de}
\affiliation{Max-Plank-Institut f\"ur Plasmaphysik, D-85748, 
    Garching, Germany}
\affiliation{\mbox{Physik-Department E28, Technische Universit\"at M\"unchen,
85748 Garching, Germany}}
\author{T. P\"utterich}
\affiliation{Max-Plank-Institut f\"ur Plasmaphysik, D-85748, 
    Garching, Germany}
\author{E. Viezzer}
\affiliation{Max-Plank-Institut f\"ur Plasmaphysik, D-85748, 
    Garching, Germany}
\affiliation{Dept. of Atomic, Molecular and Nuclear Physics, University of Seville, Avda. Reina Mercedes, 41012 Seville, Spain}
\author{R. Dux}
\affiliation{Max-Plank-Institut f\"ur Plasmaphysik, D-85748, 
    Garching, Germany}
\author{B. Geiger}
\affiliation{Max-Plank-Institut f\"ur Plasmaphysik, D-85748, 
    Garching, Germany}
\author{R. M. McDermott}
\affiliation{Max-Plank-Institut f\"ur Plasmaphysik, D-85748, 
    Garching, Germany}
\author{H. Meyer}
\affiliation{EURATOM/CCFE Fusion Association, Culham Science Centre,
Abingdon, Oxon, OX14 3DB, UK}
\author{U. Stroth}
\affiliation{Max-Plank-Institut f\"ur Plasmaphysik, D-85748, 
    Garching, Germany}
\affiliation{\mbox{Physik-Department E28, Technische Universit\"at M\"unchen,
85748 Garching, Germany}}
\author{the ASDEX Upgrade Team}
\thanks{For authors' list, see H. Zohm et al., NF 55 104010 (2015)}
\affiliation{Max-Plank-Institut f\"ur Plasmaphysik, D-85748, 
    Garching, Germany}



\date{\today}

\begin{abstract}

    In this work, a new type of high through-put Czerny-Turner
    spectrometer has been developed which allows to acquire multiple
    channels simultaneously with a repetition time on the order of
    \SI{10}{\us} at different wavelengths. The spectrometer has been
    coupled to the edge charge exchange recombination system at ASDEX
    Upgrade which has been recently refurbished with new lines of sight.
    Construction features, calibration methods, and initial measurements
    obtained with the new setup will be presented.

\end{abstract}


\maketitle 

\section{Introduction}

    Charge exchange recombination spectroscopy (CXRS) is a powerful method
    for measuring spatially resolved ion temperatures $T_{i}$, impurity
    densities $n_{\alpha}$ and velocities $v_{\alpha}$ in fusion
    plasmas~\cite{Fonck_1984}. The basic principle of CXRS exploits
    spectral lines emitted after charge exchange from neutral atoms into
    highly excited states of impurity ions. The subsequent decay of the
    excited state of the impurity ion leads to the emission of a photon at
    a specific wavelength. Thus, the ion temperature and the flow velocity
    are obtained from the Doppler width and Doppler shift and the impurity
    density from the radiance of the emission line. For magnetized fusion
    plasmas, these measurements can be combined in the radial force
    balance equation to determine the radial electric field
    $E_r$\cite{Mandl_1993}.

    In the past years, the ASDEX Upgrade tokamak (AUG) has been equipped
    with several charge exchange systems viewing the core and the edge of
    the plasma and providing temporally and spatially resolved CXRS
    profiles\cite{Viezzer_2012a,McDermott_2016}.  Each system is connected to
    a high-throughput Czerny-Turner spectrometer equipped with
    charge-coupled device (CCD) cameras. 
    So far, the minimum time integration of the systems was \SI{2.3}{ms} which is
    barely adequate to study phenomena like edge localized modes (ELMs)
    and definitely insufficient to address the evolution of $E_r$
    troughout
    the L-H transition. Several solutions have been developed, along the
    years, to reach sub-millisecond CXRS measurements in fusion
    plasmas\cite{Evensen_1995,Thomas_1997,Kobayashi_2003,Craig_2007,Uzun-Kaymak_2010}.
    Although some of them could reach a temporal resolution of even
    \SI{1}{\us}, they are often limited to one or two channels per
    spectrometer and/or to a fixed wavelength range. Here, a modified
    version of a typical Czerny-Turner spectrometer coupled with an
    electron multiplying (EM) CCD
    camera is proposed which allows fast CXRS measurements down to
    \SI{10}{\us} of time resolution for up to nine channels at different
    wavelengths.

\section{Spectrometer design}
\label{sec:spec}

    The design choices developed in this work were taken to improve the
    temporal resolution of a conventional Czerny-Turner spectrometer
    preserving the flexibility of a turnable grating, e.g.  changeable
    wavelengths, and multiple acquirable channels. Moreover, the new
    system should feature similar wavelength dispersion $\Delta\lambda/
    \mathrm{pixels}$ of approximately \SI{0.27}{\AA/pixel} at
    \SI{500}{nm}, in order to resolve velocities of \SI{1}{} to
    \SI{2}{km/s} which correspond to roughly $1/10$ of a pixel at
    \SI{500}{nm}. Such a shift is detectable given the ansatz of a
    Gaussian line emission. The resolution of lower velocities is
    prevented by uncertainties arising from atomic physics
    processes~\cite{Viezzer_2012a}.

\begin{figure}[t]
        \centering
        \includegraphics{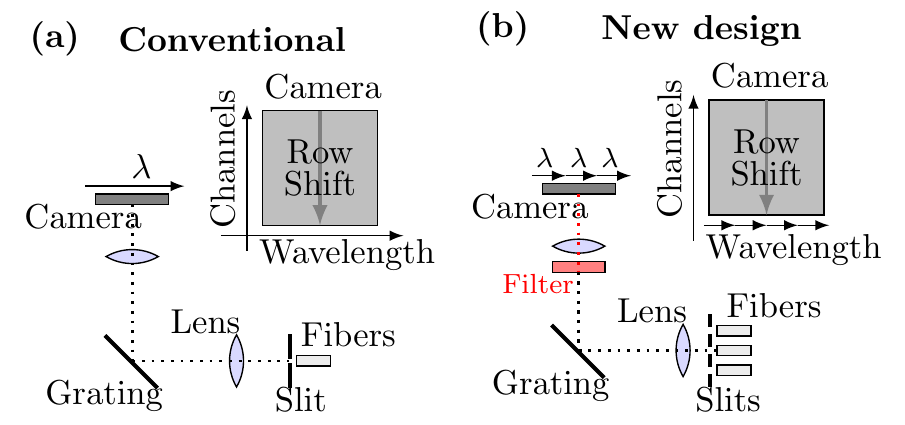}
        \caption{(a) Schematic of a conventional Czerny-Turner
            spectrometer with a single slit. (b) Schematic of the adapted
            Czerny-Turner spectrometer for fast measurements. Several
            slits are installed at the spectrometer entrance to image
            multiple channels on a smaller camera chip area. The
            interesting part of the spectra is then selected by an
            interference filter (red) positioned before the focussing
            lens.}
        \label{fig:spec}
    \end{figure}

    The principle of a conventional Czerny-Turner spectrometer is
    illustrated in figure \crefformat{figure}{#2#1{a}#3}\cref{fig:spec}.
    The light entering the entrance slit is collimated through a lens onto
    a turnable grating and re-focussed on a camera by a second lens.
    Multiple channels can be imaged simultaneously by arranging the fibers
    vertically along the entrance slit. In this way, the spectra of the
    different fibers, e.g. the wavelength axis, are not overlapping on the
    camera chip and therefore can be acquired separately. On ASDEX
    Upgrade, such spectrometers are usually equipped with a variable slit
    typically set to \SI{50}{} or \SI{100}{\um}, two objective lenses
    (Leica APO-ELMARIT-R\cite{Leica}) with each a focal length of
    \SI{280}{mm} (collimating lens) and \SI{180}{mm} (focussing lens),
    respectively, and a movable grating with \SI{2400}{grooves/mm}.  The
    lenses are chosen such that, due to the demagnification, twenty-five
    fibers with a diameter of \SI{400}{\um} can be imaged onto a Princeton
    Instruments ProEM charge coupled device (CCD) camera\cite{PI} with
    on-chip multiplication gain featuring a light sensitive area with
    $512\times 512$ \SI{16}{\um} pixels. 
    
    The temporal resolution of the system is limited by the camera
    repetition time. Its minimum $\Delta t_{\mathrm{min}}$ can be
    calculated as follows: 
    \begin{equation} 
        \Delta t_{\mathrm{min}} =
        \frac{n_{\mathrm{px,row}} \cdot n_{\mathrm{ch}}}{f_{\mathrm{ro}}} + n_{\mathrm{l}} \cdot
        v_{\mathrm{sh}} + \Delta t_{\mathrm{cam}} \label{eq:min_t}
    \end{equation} 
    where $n_{\mathrm{px,row}}$ is the number of pixel per channel,
    $n_{\mathrm{ch}}$ the number of channels, $f_{\mathrm{ro}}$ the
    read-out frequency, $n_{\mathrm{l}}$ the number of lines to shift,
    $v_{\mathrm{sh}}$ the shift speed and $\Delta t_{\mathrm{cam}}$
    eventual delays due to the camera processing such as the application
    of the pixel bias correction (PBC) or the cleaning of serial
    registers. For the old setup, e.g. for $n_{\mathrm{px,row}} = 512$,
    $n_{\mathrm{ch}} = 25$, $n_{\mathrm{l}} = 512+40(\mathrm{Mask})$,
    $v_{\mathrm{sh}}=\SI{0.45}{\us/row}$, $f_{\mathrm{ro}}=\SI{10}{MHz}$,
    the minimum acquisition time is equal to $\Delta t_{\mathrm{min,old}}
    = \SI{2.27}{ms}$. Hence, it is clear that given a certain camera, e.g.
    for fixed $f_{\mathrm{ro}},v_{\mathrm{sh}}$ and $\Delta
    t_{\mathrm{cam}}$, one needs to reduce $n_{\mathrm{px}}$ and
    $n_{\mathrm{l}}$ to improve the temporal resolution. Since the
    interesting part of the spectra only occupies between \SI{10}{} and
    \SI{20}{\%} of the entire camera chip, both $n_{\mathrm{px}}$ and
    $n_{\mathrm{l}}$ can be reduced by rearranging the imaged spectra on
    the camera chip. To do this, a re-design of the conventional
    Czerny-Turner spectrometer is required.

\begin{figure}[t]
        \centering
        \includegraphics[width=0.4\textwidth]{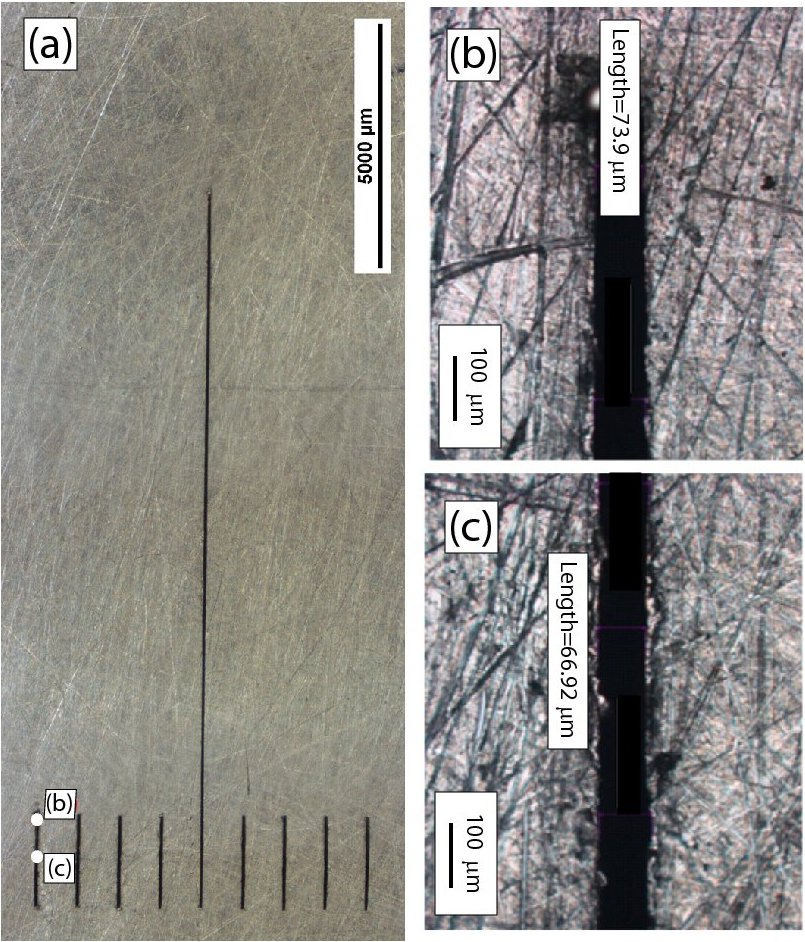}
        \caption{Entrance slits of the fast Czerny-Turner like
        spectrometer. (a) Overview of the entire entrance slits; (b,c)
        Details of bottom-left slit.}
        \label{fig:slit}
    \end{figure}

    Figure \crefformat{figure}{#2#1{b}#3}\cref{fig:spec} shows a
    schematic of the modified Czerny-Turner spectrometer developed in this
    work.
    An  interference filter (in red) has been installed between the
    grating and the focussing lens in order to select only the interesting
    part of the spectrum. This way several channels can be arranged along
    the wavelength axis ($\lambda$) perpendicular to the read-out
    direction without overlapping between neighbouring spectral lines. To
    that end, nine \SI{50}{\um} fixed slits have been installed at the
    spectrometer entrance which allow to acquire several channels
    simultaneously on a single chip row. In order to keep the full
    flexibility of using the spectrometer in the conventional way, the
    middle slit can accommodate 25 channels as for the design in figure
    \crefformat{figure}{#2#1{a}#3}\cref{fig:spec} while the other eight
    slits only two. In figure \crefformat{figure}{#2#1{a}#3}\cref{fig:slit} an
    overview of the fixed entrance slit system is shown while in figure
    \crefformat{figure}{#2#1{b,c}#3}\cref{fig:slit} the details of the
    bottom-left slit are highlighted. The parallel slits are aligned at
    one end of the long central slit in order that their projected images
    are positioned at the bottom end of the camera chip, close to the
    read-out region, with the purpose of avoiding unnecessary row shifts.

    The dispersion $\partial \lambda / \partial \mathrm{x}$, where
    $\lambda$ is the wavelength and $x$ the real space on the camera chip,
    of the Czerny-Turner spectrometer is a function  of the
    wavelength~\cite{Bell_2004} and it decreases by increasing wavelength.
    Moreover, for a certain temperature $\Delta \lambda / \lambda$ is a
    constant and therefore for longer wavelengths $\Delta \lambda$
    increases. Hence the number of possible channels which can be imaged
    without overlapping onto the camera chip depends on the wavelength.
    For instance, in case of the helium CX-line
    $\mathrm{He}^{2+}(\mathrm{n=4\rightarrow 3})$ at $\lambda =
    \SI{468.571}{nm}$ 9 channels can be simultaneously acquired while for
    the fully ionized nitrogen CX-line
    $\mathrm{N}^{7+}(\mathrm{n=9\rightarrow 8})$ positioned at $\lambda =
    \SI{566.937}{nm}$ only 5 channels can be imaged on the chip without
    overlapping.
    
    \section{CCD camera acquisition modes} 

\begin{figure}[t]
        \centering
        \includegraphics{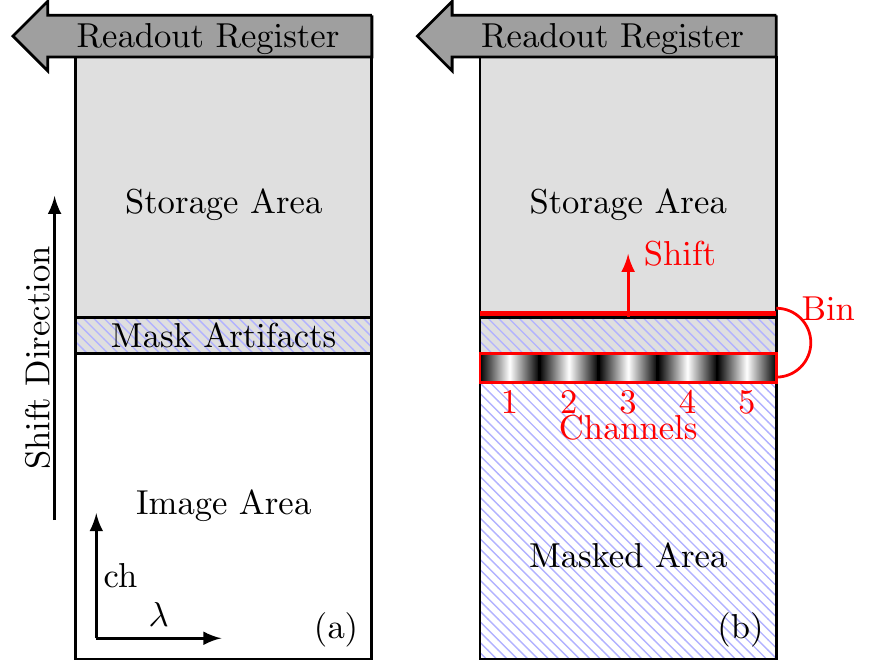}
        \caption{(a) Schematic of frame transfer CCD camera. The sensor is
            divided in two parts: the image area where the incoming
            photons are focussed and the storage area where the integrated
            charges are shifted before the read-out. (b) Working
            principle of ``Spectra Kinetics'' acquisition mode developed by
            Princeton Instruments\cite{PI}. The incoming photons are
            focussed only on a certain number of rows (highlighted in red)
            which are shifted and binned onto the first row of the storage
            area. The process is repeated until the storage region is
            filled, which is then read-out interrupting the binning and
            shifting process, i.e. leading to dead time of approximately \SI{30}{ms}.
        }
        \label{fig:camera}
    \end{figure}

    CCD cameras commonly used for high duty-cycle applications feature
    frame-transfer chips. They use a two-part sensor $512 \times 1040$ in
    which one-half of the chip
     is used as a storage area and is protected
    from light. The incoming photons are instead collected on the image
    area which, after being exposed, is rapidly shifted to the storage
    area. While the image area is exposed with light, the stored
     charges are readout. Thus, high repetition measurements are
    achieved avoiding dead times due to the chip read-out. A schematic of
    the frame-transfer CCD camera sensor is shown in figure
    \crefformat{figure}{#2#1{a}#3}\cref{fig:camera}. 
    

    Due to the re-design of the Czerny-Turner spectrometer reported in
    section \ref{sec:spec}, the acquisition mode of the CCD camera can be
    optimized to reach high repetition rates. The vertical image of a
    fiber due to the demagnification of the lenses covers roughly an area
    of 20 pixels and using the interference filter, multiple channels can
    be imaged on the same rows.  Thus, the maximum repetition time using
    the standard frame transfer acquisition mode can be reduced to
    \SI{0.71}{ms} for up to 9 channels.  A further increase of the
    temporal resolution is obtained by reducing the parallel chip size
    (this mode is often called custom chip mode).  Skipping the serial
    register cleaning and not applying the PBC, the maximum repetition
    time is \SI{0.22}{ms} for 38 vertical rows. For the Princeton
    Instruments ProEM camera, the custom chip mode can only be applied
    following precise rules in order to not interfere with the GigE Data
    Transfer requirements (specifications can be found in\cite{PI}). While
    this acquisition mode reduces the repetition time by one order of
    magnitude compared to the standard Czerny-Turner spectrometer, a
    further improvement can be obtained by exploiting the capability of
    such cameras to shift row-arrays on a sub-micro-second timescale.
    
    Figure \crefformat{figure}{#2#1{b}#3}\cref{fig:camera} shows the
    working principle of the ``Spectra Kinetics'' acquisition mode developed
    by Princeton Instruments\cite{PI}. Here, after exposure a certain number of rows
    $n_r$ (highlighted in red), are binned together at the edge of the
    image area after being exposed with light transferred into the first
    row of the storage region.  This process is extremely fast; if the
    rows, on which the light is focussed, are just at the edge storage
    area then the repetition time is equal to $\Delta t = (n_r +
    n_{\mathrm{mask}}) \cdot v_{\mathrm{sh}}$, where $n_{\mathrm{mask}}$
    is the number of rows between the image and storage regions. In case
    of the ProEM camera used in the present work, $n_{\mathrm{mask}} = 10$ and
    hence $\Delta t = (20 + 10)\, \mathrm{rows} \cdot
    \SI{0.45}{\us/rows} = \SI{13.5}{\us}$.  This can be further
    reduced down to $\approx \SI{5}{\us}$ by imaging just a part of the
    fiber onto the camera chip or by further demagnifying the fiber image.
    Note that the remaining part of the image region need to be masked
    from light to avoid interference with the acquired spectra. This
    process of binning and shifting can be repeated only until the storage
    area is filled, i.e.  528 times in case of the ProEM camera used
    here. Afterwards, the entire chip has to be read out such that a new
    burst of exposures can start. The reading out of the entire storage
    region lasts $(512 \cdot 528) \cdot f_{\mathrm{RO}} \approx
    \SI{30}{ms}$. Hence, this acquisition mode is intrinsically operating
    in a burst mode featuring dead times.

    To summarize, the best performance for continuous acquisition is
    obtained via the custom chip mode with a minimum repetition time of 
    \SI{220}{\us}. The spectra kinetic mode allows measurements
    down to \SI{5}{\us} but it operates in a burst mode with roughly
    \SI{30}{ms} of dead time.

    \section{Spectrometer calibration}
    \label{sec:calib}

    \begin{figure}[t]
        \centering
        \includegraphics{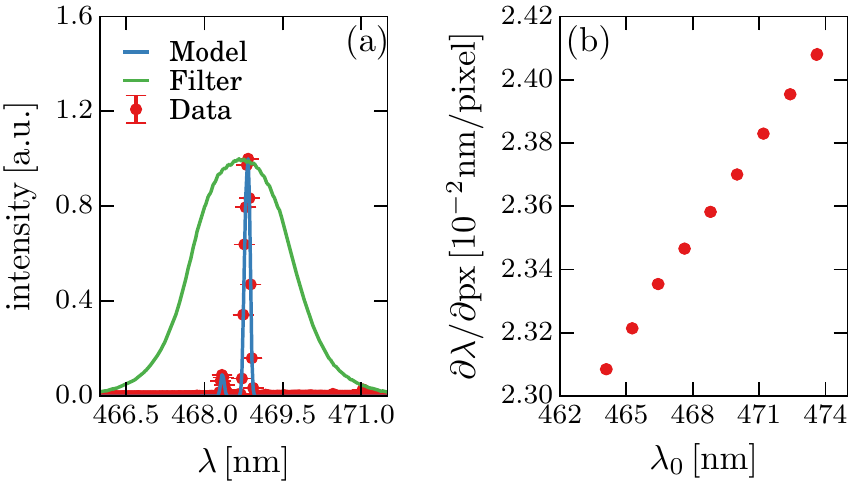}
        \caption{(a) Forward modelling (blue) of a zirconium lamp calibration
            spectra (red) in the He$^{2+}$ CX line range. The spectra has
            been deconvolved from the interference filter passband characteristic
            (green). (b) Dispersion $\partial \lambda / \partial \mathrm{px}$ in
            function of the central imaged wavelength $\lambda_0$ for the nine
            different slits.}
        \label{fig:calib}
    \end{figure}

    The interpretation of the CXRS measurements needs an accurate
    wavelength and intensity calibration.
    Wavelength calibrations are required for velocity measurements and
    the intensity calibration of the entire system (in-torus optical heads,
    connecting fibers, spectrometer and camera) is necessary for determining
    the impurity density. Moreover, the instrument function of the system has
    to be characterized to deconvolve the Doppler broadening from the
    imperfections of the spectrometer optics and to measure the ion
    temperature.

    
    During wavelength calibration the free parameters
    for the calculation of the wavelength axis~\cite{Bell_2004}, i.e. the
    central wavelength $\lambda_0$, the focal length $f_2$, the vertical
    offset of the optical axis due to eventual misalignment $\Delta h$ and
    the offset of every fiber from the central position $\Delta x_{\rm
    px}$, are determined. A forward modelling has been developed to consistently fit
    together the calibration spectra of different lamps in the wavelength
    ranges centered on the CX emissions of He$^{2+}$, B$^{5+}$, N$^{7+}$ and
    C$^{6+}$ using the previous quantities as free parameters. The
    reference wavelength for the calibration lamps have been taken from
    the NIST database~\cite{NIST}. In the forward model, parameters for
    the instrument function of every channel and for the intensity of
    every calibration line have been included to better reconstruct the
    calibration spectra.  An example of the calibration spectra  of a
    zirconium lamp and its model for one channel centered at the He$^{2+}$
    CX line is shown in
    figure~\crefformat{figure}{#2#1{a}#3}\cref{fig:calib}. Note that every
    spectrometer slit can be considered as an independent spectrometer
    with a different opening angle.  Therefore, for a fixed sine drive
    position, the imaged central wavelength and the dispersion changes for
    every vertical slit
    (figure~\crefformat{figure}{#2#1{b}#3}\cref{fig:calib}).  Moreover,
    the effect of the interference filter on the line shape has to be
    taken into account by deconvolving the calibration spectra with the
    shape of the filter passband
    (figure~\crefformat{figure}{#2#1{a}#3}\cref{fig:calib}). The accuracy
    of the calibration procedure is about \SI{2}{pm}, i.e. roughly
    \SI{1}{km/s} at \SI{468}{nm}. 

    The passband curve of an interference filters 
    exhibits temperature and angular dependences which changes with the
    wavelength~\cite{Morelli_1991}. As a result, not only the intensity of
    the observed spectral lines changes but also their shape since the
    measured line is the result of the convolution with the filter
    passband. Therefore, a dedicated calibration has been performed before
    and after the measurements while the room temperature has been
    monitored
    during the experiments. A standard calibration source (Labsphere,
    Model Unisource 1200) with a known spectral radiance has been used to
    characterize the shape and the transmission of the filters. This
    procedure guaranteed the accuracy of the intensity calibration and on the
    spectra deconvolution.
    
    \section{First measurements}

    \begin{figure}[t]
        \centering
        \includegraphics[width=0.8\textwidth]{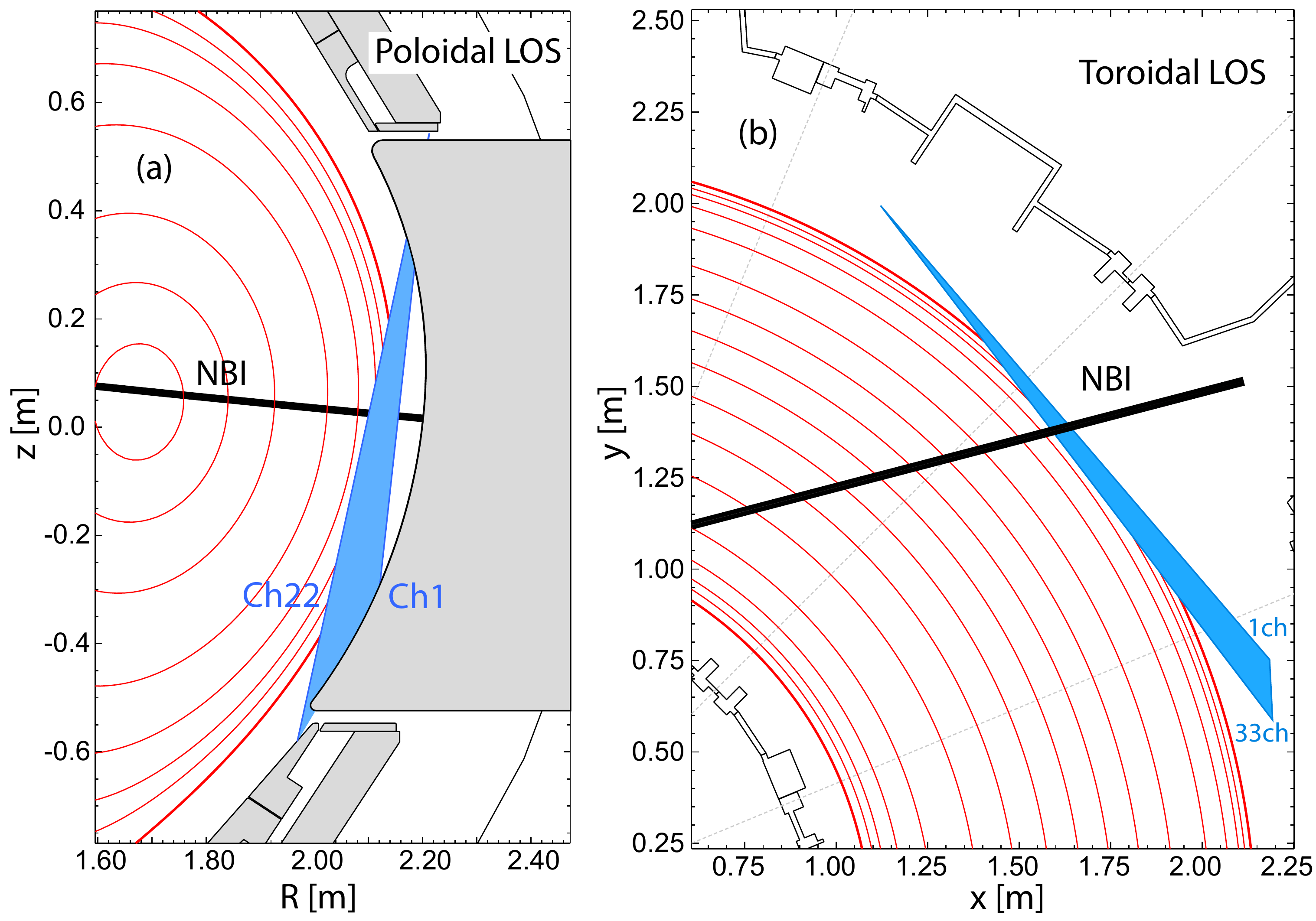}
        \caption{Line of sight of the upgraded poloidal (a) and toroidal
        (b) edge charge exchange systems at ASDEX Upgrade.}
        \label{fig:los}
    \end{figure}

    The re-designed Czerny-Turner spectrometer using the ``Spectra
    Kinetics''
    mode has successfully acquired data since the 2014 experimental
    campaign at ASDEX Upgrade. So far, only N$^{7+}$ and He$^{2+}$ CX
    emissions were considered as possible application for fast CX
    measurements because their concentration in the plasma can be
    manipulated easily by external gas puffing (``seeding'').
    This turned out to be a fundamental requirement to increase 
    the number of photons when a high repetition time is used.    The
    spectrometer has been coupled to the edge charge exchange system which
    has been upgraded with additional lines of sight to increase the
    number of radial data points in the pedestal region. The total number of
    lines of sight has been increased from 16 to 50 compared to the old
    systems~\cite{Viezzer_2012a}. Moreover, the radial coverage has been
    reduced to roughly \SI{6}{cm} in order to obtain fully resolved
    profiles of the edge ion temperature, impurity flows and density and
    hence of $E_r$, without the need of a radial plasma sweep to increase
    the radial resolution. The lines of sight of the refurbished poloidal
    and toroidal systems are shown in figure
    \crefformat{figure}{#2#1{a}#3}\cref{fig:los} and
    \crefformat{figure}{#2#1{b}#3}\cref{fig:los}. The radial resolution of
    the toroidal LOS is around \SI{3}{mm} at the very edge of the plasma
    while it increases towards the inner part of the plasma. Since the
    poloidal curvature of the plasma is stronger than the toroidal one,
    the resolution of the poloidal system is reduced to about \SI{5}{mm}.
    Good signal to noise ratio measurements have been obtained for a time
    resolutions down to \SI{50}{\us}.
    
    \begin{figure}[t]
        \centering
        \includegraphics{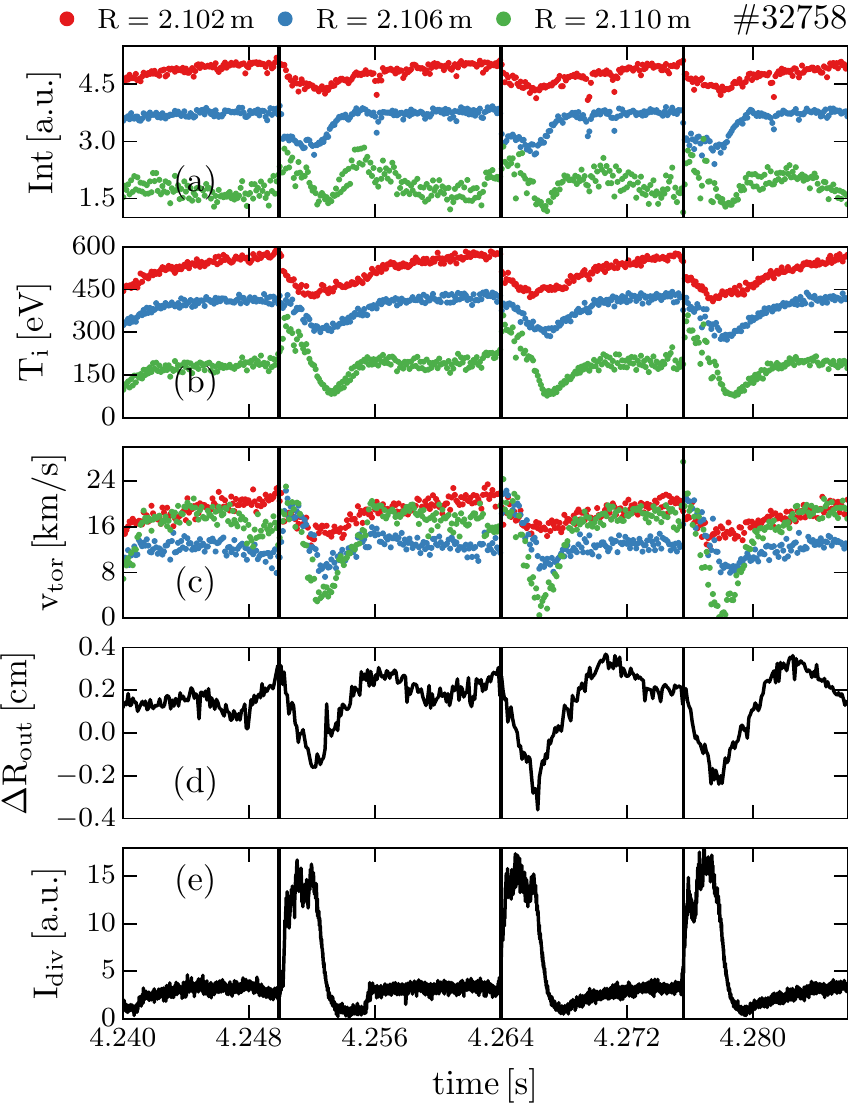}
        \caption{Fast charge-exchange measurements of the
            \mbox{He$^{2+}(n=4\rightarrow 3)$} line emission during
            edge-localized modes in a helium plasma with a temporal
            resolution of \SI{50}{\us}. (a) Line intensity; (b) Ion
            temperature; (c) Toroidal velocity; (d) Radial excursion of
            the plasma; (e) Divertor shunt current. The different colors
            indicate different radial positions.
        }
        \label{fig:ELM}
    \end{figure}

    As an initial check of the time resolution of the system, an ELMy H-mode
    discharge has been examined. In figure \ref{fig:ELM}, time traces of the
    line intensity (a), the ion temperature (b) and the toroidal velocity
    (c) derived from the He$^{2+}(n=4\rightarrow 3)$ CX line are shown.
    Here, the different colors indicate different radial positions.  The
    measurements were obtained in a helium discharge and the spectra were
    acquired with a temporal resolution of \SI{50}{\us}.  The profile
    information is
    compared to the radial excursion of the plasma (fig.
    \crefformat{figure}{#2#1{d}#3}\cref{fig:ELM}) and with the divertor
    shunt currents (fig. \crefformat{figure}{#2#1{e}#3}\cref{fig:ELM})
    which is often used as an ELM monitor signal. Note that the data
    shown in figures \crefformat{figure}{#2#1{a-}#1{c}#3}\cref{fig:ELM}
    are obtained by fitting single channels. Hence, the time-traces are
    shown at a fixed radial position while the plasma position changes
    slightly. In
    particularly, during the ELM crash the plasma position can change by
    $\Delta R_{\mathrm{out}} = \SI{5}{mm}$ which is larger than the
    resolution of the system and therefore this effect needs to be
    accounted for in the flux surface mapping. The measured time traces show
    low statistical noise. The relative standard variation $\sigma_r$ is
    defined as the ratio of the standard deviation to the mean is
    calculated by binning a time window of \SI{1}{ms}, i.e. 20
    measurements, during a relatively constant plasma phase before the ELM
    onset. The values are roughly equal to: $\sigma_r (v_\mathrm{tor}) = \SI{4}{\%}$
    for the toroidal velocity, $\sigma_r(T_i) = \SI{1}{\%}$ for the ion
    temperature and $\sigma_r(\mathrm{inte}) = \SI{3}{\%}$ for the line
    intensity. The accuracy of the measured profiles combined with the
    high temporal resolution open the possibility at ASDEX Upgrade to
    characterize in detail the evolution of the ion temperature and
    impurity flow profiles during ELMs and to compare the behaviour to the
    electron profiles similarly as in Refs.~ \cite{Viezzer_2013} and~
    \cite{Wade_2005a} but with better time resolution. Moreover, due to the
    low noise and by increasing the time resolution and
    the averaging time window it might be possible to monitor fluctuations of the ion
    temperature $\tilde{T_i}/\bar{T_i}$ and of the impurity flows
    $\tilde{v}/\bar{v}$~\cite{Evensen_1995,Uzun-Kaymak_2012}.

\begin{figure}[t]
        \centering
        \includegraphics{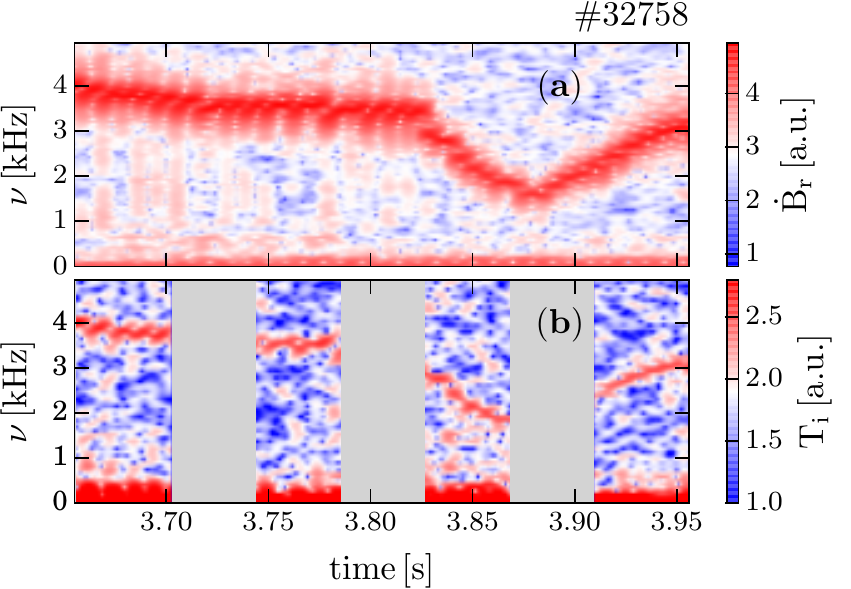}
        \caption{(a) Spectrogram of $\dot B_r$ measured at the low field
            side midplane. (b) Spectrogram of $T_i$ at $\rho_{\rm
            pol}=0.995$. The gray areas in figure (b) indicate the dead
            time in between two burst of the fast acquisition system.}
        \label{fig:spectrogram}
    \end{figure}

    With the new setup, it has also been possible to detect
    magnetohydrodynamic activities at low frequency. An example is shown
    in figure \ref{fig:spectrogram} where the spectrogram of a ballooning
    coil measuring $\dot B_r$ at the low field side mid-plane of the
    plasma (fig. \crefformat{figure}{#2#1{a}#3}\cref{fig:spectrogram}) is
    compared to the spectrogram of $T_i$ measured at $\rho_{\rm pol} =
    0.995$ (fig. \crefformat{figure}{#2#1{b}#3}\cref{fig:spectrogram}).
    The gray shaded areas in figure
    \crefformat{figure}{#2#1{b}#3}\cref{fig:spectrogram} indicate the dead
    time in-between two bursts.  Both measurements see a distinct mode at
    around \SI{4}{kHz} (highlighted in red) changing its frequency at
    around \SI{3.85}{s}. The mode is a global movement of the plasma which
    at a fixed radial position results in a fluctuation of the plasma
    parameters such as $T_i$. The detected fluctuations in $T_i$ induced
    by the mode are of the order of \SI{2}{\%} and are strongest in the
    steep gradient region where a small displacement of the plasma induces
    the largest variation of the temperature.
    
    Similar investigations have been attempted in N-seeded discharges in
    order to measure the nitrogen CX lines
    $\mathrm{N}^{7+}(\mathrm{n=9\rightarrow 8})$ with sub-ms time
    resolution.   However, the nitrogen concentration could not be
    increased above roughly \SI{3}{\%} without incurring into a
    disruption~\cite{Bernert_2016}. Hence, to reach good signal to noise
    ratio, the integration time could not be decreased below
    \SI{200}{\us}. In this case, a brighter spectrometer or in-torus
    optics are necessary to further optimize the temporal resolution. 

    \section{Summary and future plans}

    In summary, a multi-channel spectrometer with about $\SI{10}{\us}$ of
    time resolution dedicated to edge charge exchange recombination
    spectroscopy has been developed at ASDEX Upgrade to characterize the
    fast evolution of ion temperature, impurity density and flows. A
    multi-slit design equipped with an interference filter allows for the
    simultaneous acquisition of multiple channel. The spectrometer has
    been coupled to the edge poloidal and toroidal charge exchange systems
    which have been refurbished with 50 new lines of sight covering a
    radial range of about \SI{6}{cm}. The new setup offers the possibility
    to measure detailed profiles during fast events without the necessity
    of a radial sweep of the plasma.

    Good signal to noise ratio has been obtained for an integration time of
    \SI{50}{\us} during edge localized modes measuring the He$^{2+}$
    charge exchange emission. The measurements achieved a statistical
    noise of a few percent. Hence, it is expected that using the highest
    temporal resolution (\SI{10}{\us}), low frequency ion temperature and
    impurity flow fluctuations might be detectable.
    
    A larger CCD sensor would allow the simultaneous acquisition of more
    channels while the read-out time, i.e. the acquisition dead time,
    would be increased.  More spectrometer channels can be employed by
    increasing the spectrometer dispersion but this might influence the
    accuracy of the velocity measurements.  While the working principle of
    the modified Czerny-Turner spectrometer has been successfully tested,
    different setups need to be investigated to relatively optimize the
    number of acquirable channels, the acquisition dead time and the
    measurement accuracy. The optimum solution might depend on the
    investigated phenomena. 
    



\bibliography{rsi.bib}

\end{document}